\documentclass[%
 reprint,
 amsmath,amssymb,
 aps,pra
]{revtex4-2}
\usepackage{graphicx}
\usepackage{amsmath}
\usepackage{nicefrac}
\usepackage[hidelinks]{hyperref}
\usepackage{braket}
\usepackage{amssymb}
\usepackage{siunitx}
\DeclareSIUnit\gauss{G}
\usepackage{xspace}
\usepackage{mathtools}

\newcommand{\swap}{\ifmmode\text{\textsc{swap}}\else\textsc{swap}\fi\xspace}
\newcommand{\swapsq}{\ifmmode\text{\textsc{swap}}^2\else$\text{\textsc{swap}}^2$\fi\xspace}
\newcommand{\swapd}{\ifmmode\sqrt{\text{\sc swap}}^{\dagger}\else$\sqrt{\text{\sc swap}}^{\dagger}$\fi\xspace}
\newcommand{\swapa}{\ifmmode(\text{\sc swap})^{\alpha}\else$(\text{\sc swap})^{\alpha}$\fi\xspace}
\newcommand{\sswap}{\ifmmode\sqrt{\text{\sc swap}}\else$\sqrt{\text{\sc swap}}$\fi\xspace}

\begin{document}

\title{
Protected quantum gates using qubit doublons in dynamical optical lattices 
}
\author{Yann Kiefer}
\thanks{These authors contributed equally}
\author{Zijie Zhu}
\thanks{These authors contributed equally}
\
\author{Lars Fischer}
\author{Samuel Jele}
\author{Marius Gächter}
\author{Giacomo Bisson}
\author{Konrad Viebahn}
\email[]{viebahnk@phys.ethz.ch}
\author{Tilman Esslinger}
\affiliation{Institute for Quantum Electronics \& Quantum Center, ETH Zurich, 8093 Zurich, Switzerland}



\maketitle

\textbf{
Quantum computing represents a central challenge in modern science.
Neutral atoms in optical lattices have emerged as a leading computing platform, with collisional gates offering a stable mechanism for quantum logic~\cite{jaksch_entanglement_1999,brennen_quantum_1999,sorensen_spin-spin_1999,calarco_quantum_2000,hayes_quantum_2007,daley_quantum_2008,weitenberg_quantum_2011,nemirovsky_fast_2021,zhou_scheme_2022,singh_optimizing_2025}.
However, previous experiments have treated ultracold collisions as a dynamically fine-tuned process~\cite{mandel_controlled_2003,anderlini_controlled_2007,kaufman_entangling_2015,folling_direct_2007,trotzky_controlling_2010,zhu_splitting_2025,chalopin_optical_2025,yang_cooling_2020,zhang_scalable_2023,bojovic_high-fidelity_2025,murmann_two_2015,hartke_quantum_2022}, which obscures the underlying quantum-geometry and -statistics crucial for realising intrinsically robust operations.
Here, we propose and experimentally demonstrate a purely geometric two-qubit \swap gate by transiently populating qubit doublon states of fermionic atoms in a dynamical optical lattice.
The presence of these doublon states, together with fermionic exchange anti-symmetry, enables a two-particle quantum holonomy—a geometric evolution where dynamical phases are absent~\cite{sjoqvist_geometric_2015}.
This yields a gate mechanism that is intrinsically protected against fluctuations and inhomogeneities of the confining potentials.
The resilience of the gate is further reinforced by time-reversal and chiral symmetries of the Hamiltonian.
We experimentally validate this exceptional protection, achieving a loss-corrected amplitude fidelity of $99.91(7) \%$ measured across the entire system consisting of more than $17'000$ atom pairs.
When combined with recently developed topological pumping methods for atom transport~\cite{zhu_splitting_2025}, our results pave the way for large-scale, highly connected quantum processors.
This work introduces a new paradigm for quantum logic which transforms fundamental symmetries, including quantum statistics, into a powerful resource for fault-tolerant computation.
}

\begin{figure}[h!]
    \includegraphics[width=0.48\textwidth]{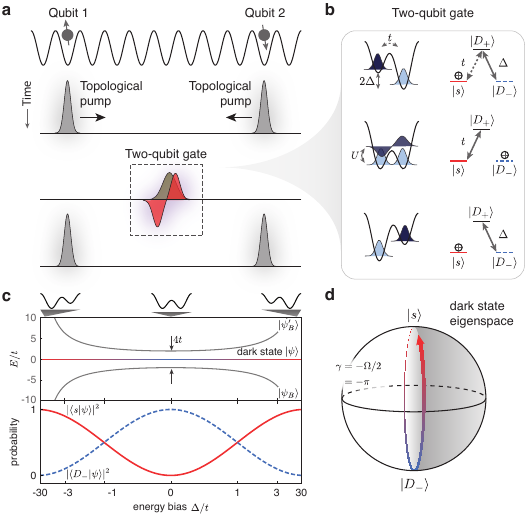}
         \caption{\textbf{Two-qubit gates realised by populating qubit doublon states.} (a) Fermionic qubits are initially trapped on individual lattice sites, isolated from one another.
         To initiate two-qubit gates, the spatial wavefunction of both qubits are brought to overlap, e.g.~via topological pumping~\cite{walter_quantization_2023,zhu_splitting_2025}, temporarily forming qubit doublons. Wavefunction overlap brings out the indistinguishability of both qubits and thereby exchange statistics and quantum-geometrical effects become relevant, leading to a robust two-qubit gate mechanism (b).
         The double-well potential (left) with tuneable  bias \(\Delta\), tunnelling \(t\), and Hubbard $U$ describes two distinct spatial orbitals (purple, blue) which are involved in the gate protocol. (right) $\Lambda$-system representation of the state transfer from $\ket{s}$ to $\ket{D_-}$ and vice versa within the singlet subspace $\mathcal{S}$ (see main text). (c, top) Energy spectrum of $\mathcal{S}$ as a function of bias $[\Delta/t = -\cot(\theta/2)]$ following the gate sequence used in the experiment in which $\Delta/t$ is swept from large negative to large positive values during the gate time $\tau$ for Hubbard $U=0$. The triplet states $\mathcal{T}$ always remain at zero energy. (c, bottom) State composition of the dark state $\ket{\psi}$ during the gate operation. In the dimerised configuration ($\Delta/t=0$), the dark state $\ket{\psi}$ equals $\ket{D_-}$, whereas in the staggered ($\lvert \Delta\rvert  \gg t$) configurations  $\ket{\psi}=\pm\ket{s}$. (d) Trajectory of  dark state $\ket{\psi}$ during one gate operation in the dark-state eigenspace enclosing the solid angle $\Omega=2\pi$. The geometric phase is $\gamma=-\Omega/2=-\pi$.
        }
    \label{fig:1}
\end{figure}

We propose and demonstrate a geometric \swap mechanism that relies on populating qubit doublons, i.e.~two qubits occupying the same orbital, and follows a dark state during gate operation.
While the dark state picks up a geometric phase, all other states taking part in the gate evolution (in our case spin triplets) are decoupled from one another due to fermionic anti-symmetry.
The doublon states enlarge the relevant Hilbert space and allow tracing a geometric loop during gate operation.
Crucially, due to the dark-state nature of the process, all dynamical phases are zero by construction, forming a quantum holonomy, i.e.~a quantum evolution in which dynamical effects vanish.
Previous experimental realisations of quantum holonomies have been primarily limited to the single-particle regime, such as the tripod system~\cite{sjoqvist_geometric_2015,duan_geometric_2001}, while we describe an all-geometric two-particle evolution here.
Since the spatial wavefunctions of the qubits are made to overlap (qubit doublon), it is important to operate in the quantum-degenerate regime in which the exchange statistics of indistinguishable qubits are physically relevant, rather than being treated as thermal particles.
Both fermionic and bosonic statistics give rise to a well-defined geometric phase (Methods - \ref{subsec:bosons}), but we focus on the fermionic case for the remainder of the paper.
The geometric phase is further protected by time-reversal and chiral symmetry, making it robust against fluctuations in all experimental control parameters of the model.

Our implementation operates with fermionic atoms in a dynamical optical lattice.
In the idle state, a deep lattice suppresses quantum tunnelling and ensures that qubits remain spatially localised and decoupled from one another.
Protected gate operations then result from controllably overlapping the qubits' spatial wavefunctions (Fig.~\ref{fig:1}a), forming qubit doublons in the process, and realising a geometric \swap.
Importantly, the geometric \swap gate is insensitive to lattice inhomogeneities, which is important for scaling up to even larger system sizes.
When adding interactions, we further realise entangling \swapa gates which are likewise robust against fluctuations in the tunnel coupling of the lattice, despite not operating in a single-well harmonic oscillator (as opposed to refs.~\cite{anderlini_controlled_2007,kaufman_entangling_2015}).

\label{sec:1}

\section{Qubit doublons and geometric swap}

The double-well potential is a minimal model to describe protected two-qubit operations as it captures both spatially isolated qubits, as well as delocalised qubits with doublon states.
Specifically, we consider a Fermi-Hubbard model on a dimer, which is broadly applicable to cold atoms~\cite{murmann_two_2015,desbuquois_controlling_2017} and semiconductor spin qubits~\cite{burkard_semiconductor_2023,hu_gate_2002,kaplan_spin_2004}, highlighting the generality of our approach.
The model is fully parametrised by the tunnelling energy $t$, energy bias $\Delta$, and Hubbard interaction $U$ (Fig.~\ref{fig:1}b).
In the basis $\mathcal{H} = \{
\ket{\uparrow,\uparrow},\ket{\uparrow\downarrow,0}, \ket{\uparrow,\downarrow}, \ket{\downarrow,\uparrow}, \ket{0,\uparrow\downarrow},\ket{\downarrow,\downarrow}\}$ the Hamiltonian takes the form
\begin{eqnarray}\label{eq:Hfull}
\hat{H}_\text{full} = \begin{pmatrix}
0 & 0&0 &0 &0 &0 \\
0& U + 2\Delta & -t & t & 0 &0 \\
0& -t & 0 & 0 & -t & 0\\
0& t & 0 & 0 & t & 0\\
0& 0 & -t & t & U - 2\Delta &0\\
0&0 &0 &0 &0 & 0
\end{pmatrix}\,.
\end{eqnarray}
We choose site-ordered normal ordering and $\ket{\uparrow,\downarrow}$ denotes a state with a spin-$\uparrow$ fermion on the left site and a spin-$\downarrow$ on the right.
While the computational space is restricted to
$\mathcal{C} = \{\ket{\uparrow,\uparrow}, \ket{\uparrow,\downarrow}, \ket{\downarrow,\uparrow}, \ket{\downarrow,\downarrow}\}$, it is crucial to include the doublon states, $\ket{\uparrow\downarrow,0}$ and $\ket{0,\uparrow\downarrow}$, to realise geometric and fast dynamical two-qubit gates in $\mathcal{C}$.
To simplify their analytical description, we perform a basis transformation from $\mathcal{H}$ to eigenstates of the spin operator~\cite{auerbach_interacting_1994}, which can be grouped to triplet states  $ \mathcal{T} = \{\ket{t_+} =\ket{\uparrow,\uparrow},\ket{t_0} = (\ket{\uparrow,\downarrow}+ \ket{\downarrow,\uparrow})/\sqrt{2},\ket{t_-} =\ket{\downarrow,\downarrow}\}$
and singlet states 
\begin{eqnarray}\label{eqn:singlet-basis}
\mathcal{S} = \{\ket{D_+} &=& (\ket{\uparrow\downarrow,0} + \ket{0,\uparrow\downarrow})/\sqrt{2},\\
\nonumber\ket{D_-} &=& (\ket{\uparrow\downarrow,0} - \ket{0,\uparrow\downarrow})/\sqrt{2},\\
\nonumber\ket{s} &=& (\ket{\uparrow,\downarrow}- \ket{\downarrow,\uparrow})/\sqrt{2}\}~.
\end{eqnarray}
States within $\mathcal{S}$ are coupled by
\begin{equation}\label{eq:Hsinglet}
\hat{H}_\mathcal{S} = \begin{pmatrix}
U & 2\Delta & -2t \\
2\Delta & U & 0 \\
-2t & 0 & 0
\end{pmatrix}~,
\end{equation}
while states in $\mathcal{T}$ have zero energy and remain completely decoupled throughout, due to fermionic anti-symmetrisation, serving as phase reference for robust quantum gate evolution.
We construct partial exchange gates -- $(\text{\textsc{swap}})^\alpha$ -- by imparting controlled dynamical ($\delta$) and/or geometric ($\gamma$) phases on the $\ket{s}$ state, i.e.
$\ket{s} \longrightarrow e^{i\varphi} \ket{s}$.
A key result of our work is the realisation of a purely geometric phase $\varphi = \gamma = -\pi$ via a two-particle holonomy for constructing robust \swap gates in $\mathcal{C}$.
Additionally, a total acquired phase of $\varphi = \pi/2$ at finite $U$ corresponds to an entangling \sswap gate (Section~\ref{sec:direct-exchange}b).

The geometric origin of the \swap gate can be understood in analogy to the three-level atom in the $\Lambda$-configuration (Fig.~\ref{fig:1}b, refs.~\cite{duan_geometric_2001,vitanov_stimulated_2017}), whose Hamiltonian is equivalent to $\hat{H}_\mathcal{S}$ at $U = 0$.
The spectrum (Fig.~\ref{fig:1}c, top) features a zero-energy dark state which arises as a coherent superposition of $\ket{s}$ and $\ket{D_-}$, and remains uncoupled from $\ket{D_+}$.
We write the dark state state as 
\begin{equation}\label{eq:dark_state_total}
    \ket{\psi} = \cos\left(\frac{\theta}{2}\right)\ket{s}+\sin\left(\frac{\theta}{2}\right)\ket{D_-}~.
\end{equation}
The mixing angle $\theta \in [0,2\pi]$, parametrised as
\begin{equation}
    \cot \left(\frac{\theta}{2}\right) = -\frac{\Delta}{t}\, ,
\end{equation}
defines how $\ket{\psi}$ changes between $\ket{s}$ and $\ket{D_-}$, as  shown in Fig.~\ref{fig:1}c (bottom).

Quantum gate operation proceeds as follows.
Starting from $\theta \simeq 0$ (large negative energy bias and negligible tunnelling, Fig.~\ref{fig:1}b, first row), the dark state equals $\ket{s}$ and qubits are decoupled from one another on their respective lattice sites.
Subsequently, we adiabatically sweep the energy bias from large negative to large positive values, thereby changing $\theta$ from $0$ to $2\pi$ (Fig.~\ref{fig:1}b,c).
Crucially, the instantaneous Hamiltonian $\hat{H}_\mathcal{S}$ preserves time-reversal symmetry, all couplings being real-valued, leading to real-valued state vectors during gate operation (in the adiabatic limit), which further constrains the quantum trajectory to a great circle on the Bloch sphere (Fig.~\ref{fig:1}d, Methods - \ref{subsec:ExGate_robust}).
All states taking part in the gate operation $\{\mathcal{T}, \ket{\psi}\}$ remain at zero energy throughout (for $U = 0$),  protected by a chiral symmetry in $\hat{H}_\mathcal{S}$ (Methods - \ref{subsec:ExGate_robust}).
Therefore, the process exhibits a two-particle quantum holonomy and any phase acquired by $\ket{s}$ is purely geometric.
Specifically, the geometric Aharonov-Anandan phase~\cite{aharonov_phase_1987} equals half the solid angle subtended by $\psi(\tau)$, i.e. $\gamma = -\Omega/2 = -\pi$ (Fig.~\ref{fig:1}d and Methods - \ref{subsec:Geometric_phase}). 

Although we describe an adiabatic quantum evolution, the gate need not be slow:
The dark state is gapped from the `bright' states $\ket{\psi_B}$ and $\ket{\psi_B'}$ by at least $2t$ (Fig.~\ref{fig:1}c). The adiabatic criterion $\partial \theta/\partial\tau \ll 2t$ can easily be fulfilled in our experiments by making $t$ large, leading to sub-millisecond timescales.
To further increase gate speed, the double-well potential can be fully merged, leading to $2t \rightarrow \hbar \omega$, where $\omega$ is the trap frequency of the merged single-well.

The quantum holonomy results from deliberate population of qubit doublon states, which were previously considered unwanted leakage in previous gate realisations~\cite{hu_gate_2002}.
The value of the geometric phase $\gamma$ is fully constrained by solid angle $\Omega$ and enables a remarkable robustness of quantum gate operation against experimental control parameters.
While fluctuations and inhomogeneities in the confining lattice potential may affect the rate at which the trajectory on the Bloch sphere is traversed, they do not alter its shape (Fig.~\ref{fig:1}d).
This geometric robustness is reminiscent of the quantised charge transport observed in topological pumps and our gate scheme can readily integrated into a topological pump architecture with large-scale and non-local qubit connections~\cite{zhu_splitting_2025}.

\section{Experimental realisation of the geometric SWAP gate}

\begin{figure}
    \includegraphics[width=0.48\textwidth]{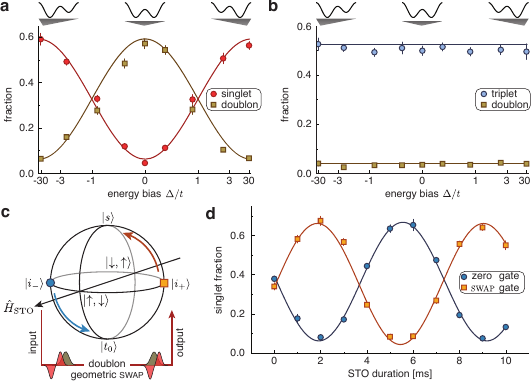}
         \caption{\textbf{Experimental demonstration of the geometric SWAP gate.} (a) Measurement of the singlet (red circles) and doublon (brown squares) fraction as a function of the bias $\Delta/t$ during the gate operation. The measurment shows the occurence of the doublon state $\ket{D_-}$ in the dimerised configuration $\Delta/t=0$ of the lattice. (b) Measurement of the triplet (blue circles) and doublon (brown squares) fraction as a function of the ratio $\Delta/t$ showing that the triplet state $\ket{t_0}$ does not evolve during the gate. The solid lines denote the theoretical evolution with no free parameters (a and b). (c) Two-particle $\ket{s}\,\text{-}\,\ket{t_0}$ Bloch-sphere  including the two-particle states $\ket{i_-},\ket{i_+}$ relevant for state preparation and detection. Starting from atomic singlets $\ket{s}$, we rotate $\ket{s}$ to $\ket{i_-}$ on the equator using a quarter singlet-triplet oscillation (STO), which is also applied for detection. The geometric \swap happens outside the $\ket{s}\,\text{-}\,\ket{t_0}$ sphere. In (d), we show the singlet fraction as a function of the STO duration after applying one \swap gate (orange squares) and compare it to the scenario in absence of any gate (blue circles). The data in (a, b, d) were collected from $10$ experimental realisations, with error bars representing the standard error.}
    \label{fig:2}
\end{figure}
We experimentally demonstrate the geometric \swap gate using adiabatic evolution in a dynamical superlattice. The experiment utilises $5.8(2)\times10^{4}$ fermionic $^{40}$K atoms, of which $60\,\text{-}\,70$\% (more than $17'000$ pairs) are prepared in the singlet state $\ket{s}$, within the double-well Fermi-Hubbard system (see Methods - \ref{subsec:statePrep}).
The qubit is defined as $\{\ket{\downarrow} = \ket{F = 9/2, m_F = -9/2}, \ket{\uparrow} = \ket{F = 9/2, m_F = -7/2}\}$.
The gate is realised by sweeping the mixing angle $\theta$ from 0 to $2\pi$, which corresponds to a full ramp of $\Delta/t$ between the staggered ($\Delta\approx \SI{4}{\kilo\hertz},t\approx\SI{0.1}{\kilo\hertz}$) and the dimerised ($\Delta=\SI{0}{\kilo\hertz}, t\approx\SI{3}{\kilo\hertz}$) configurations of the lattice (see Methods -~\ref{subsec:SLPotential}).
The dark state $\ket{\psi}$ changes between the $\ket{s}$ and the doublon state $\ket{D_-}$ during gate evolution (Eq.~\ref{eq:dark_state_total}), while all three triplet states $\mathcal{T}$ remain decoupled.
We measure the state composition during a single gate operation by initializing the state in either the singlet $\ket{s}$ or the triplet $\ket{t_0}$ state and sweeping the energy bias $\Delta/t$.
The measurements in Figure \ref{fig:2} demonstrate how the singlet evolves to $\ket{D_-}$ and back (Fig.~\ref{fig:2}a), while the triplet fraction shows no change in dependence of the bias $\Delta/t$ (Fig.~\ref{fig:2}b), confirming that the state $\ket{t_0}$ is decoupled.
To verify the action of the geometric \swap gate and, in particular, the acquisition of the geometric phase $\gamma=-\pi$, we conduct an experiment in which the system is initialised in the $\ket{i_-}$ state (see Methods -  \ref{subsec:statePrep}). Following the state preparation, we apply the \swap gate with a total duration $\tau.$ Unless otherwise specified, the gate duration is always fixed to $\tau=\SI{750}{\mu\s}$, realising a sub-millisecond two-qubit gate. This operation transforms the input state,
\begin{equation*}
    \ket{i_-}=(\ket{t_0}-i\ket{s})/\sqrt{2}\xrightarrow{\swap}\ket{i_+}=(\ket{t_0}+i\ket{s})/\sqrt{2},
\end{equation*}
as illustrated by the red arrow in Figure \ref{fig:2}c on the two-particle Bloch sphere spanned by $\{ \ket{s},\ket{t_0 }\}$. The acquired phase $\gamma=-\pi$ between the singlet state $\ket{s}$ and the triplet state $\ket{t_0}$ is of purely geometric origin as all participating states have zero energy and therefore the dynamical phase $\delta$ is fixed to zero.
The successful implementation of the geometric \swap gate is verified by inducing coherent singlet-triplet oscillations (STOs) via application of a magnetic gradient of variable duration (refs.~\cite{trotzky_controlling_2010,greif_short-range_2013} and Methods - B).
The orientation of the STO-Hamiltonian $\hat{H}_{\text{STO}}$ causes the $\ket{i_{\pm}}$ states to have initial oscillation phases differing by $\pi$, as illustrated by the colored arrows in Figure \ref{fig:2}c. 
In Figure \ref{fig:2}d, we plot the singlet fraction as a function of the STO duration, comparing the geometric \swap-gate implementation to the scenario in absence of any gate. The observed $\pi$-phase shift, together with unchanged oscillation amplitude, confirms the realisation of a geometric \swap gate applied to the input state $\ket{i_-}$.

\textit{Geometric swap gate fidelity}.
\begin{figure}[t!]
    \includegraphics[width=0.48\textwidth]{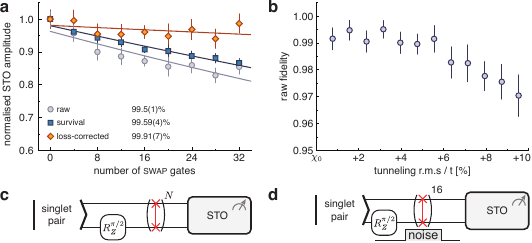}
         \caption{\textbf{Fidelity and robustness of the geometric \swap gate.} (a) Measurement of the normalised STO amplitude of the geometric \swap gate as a a function of the number of applied gates. We plot the raw amplitude (grey circles) and the survival amplitude (blue squares). The loss-corrected STO amplitude is obtained by conditioning the raw data on the survival of atoms (orange diamonds). The data points and error bars are obtained by fitting a sinusoidal function to the STOs of $11\,\text{-}\,15$ experimental realisations (Methods). The fidelity is obtained extracting the time constant of an exponential decay function and its error is given by the fit uncertainty (see Methods - \ref{subsec:STO_Fidelity_correction}).  (b) We plot the raw fidelity as a function of added tunnelling root-mean-square noise on the lattice potential. We apply noise of variable strength at a bandwidth of $2$kHz for the duration of 16 gates and subsequently measure the atomic singlet fraction. The projected fidelity is obtained from the singlet fraction based on $9\,\text{-}\,10$ experimental realisations, with error bars showing the standard error (see Methods - \ref{subsec:STO_Fidelity_correction}). (c,d) Circuit representation for the measurement protocols used in a and b, respectively.  
        }
    \label{fig:3}
\end{figure}
We assess the performance of the the \swap gate by measuring the gate fidelity.
As before, we initialise the atom pairs in the $\ket{i_-}$ state. We proceed to apply a variable number of \swap gates by periodically cycling between the negative and positive staggered configurations of the lattice potential (see Fig.~\ref{fig:3}c and Methods - \ref{subsec:expSeq_ReadOut}). Subsequently, we measure the STO amplitude and use it to quantify the fidelity of the gate (see Methods - \ref{subsec:STO_Fidelity_correction}). In Figure \ref{fig:3}a, we plot the normalised STO amplitude as a function of the number of \swap gates. The fidelity of repeated two-particle \swap gate applications is determined by fitting an exponential decay function and using the extracted decay constant. The raw fidelity of the gate operation is $99.5(1)\%$. To correct for dispersive losses during the application of the gates, we can measure the survival fidelity $99.59(4)\%$ and calculate the corrected fidelity conditioned on the survival of the two-particle state $99.91(7)\%$.
These measured fidelities represent a direct average over more than $17'000$ pairs, including system-wide lattice inhomogeneity and disorder.

\textit{Robustness against tunnelling noise}. Geometric phases are expected to be inherently protected against perturbations in the control parameters. Therefore, we introduce white noise intensity modulation of variable amplitude at a bandwidth of $\SI{2}{\kilo\hertz}$ onto the optical lattice potential $V_X$ (see Fig. \ref{fig:3}d). The modulation bandwidth is chosen to be larger than the gate cycle frequency but small compared to single-particle energy gaps and the intensity modulation predominantly adds noise to the tunnelling coupling $t$ of the double well.
The noise contribution given by the other lattice beams is summarised in a single parameter $\chi_0$. We apply the noise for the duration of 16 gates and measure the STO amplitude, from which we infer the raw fidelity of the geometric \swap gate (see Methods - \ref{subsec:STO_Fidelity_correction}).
In Figure \ref{fig:3}b, we plot the raw fidelity as a function of the added noise amplitude, which is expressed as a percentage of the tunneling $t$.
We observe a distinct plateau of the fidelity extending to up to $\SI{5}{\percent}$ of added tunnelling noise, indicating robustness of the geometric \swap gate against significant variations of lattice control parameters. This geometric protection is effective as long as the system evolves adiabatically within the dark-state manifold, which is protected by an energy gap of at least $2t$ from other states. The fidelity drops for noise amplitudes above 5\% because high-frequency components of the noise can drive non-adiabatic transitions across this gap, breaking the protection and leading to errors. The robustness window is therefore defined by noise frequencies well below this characteristic energy gap (see Methods - \ref{subsec:noise_contributions}).

\section{Robust dynamical phases in the direct exchange regime}\label{sec:direct-exchange}
\begin{figure}[t]
    \includegraphics[width=0.48\textwidth]{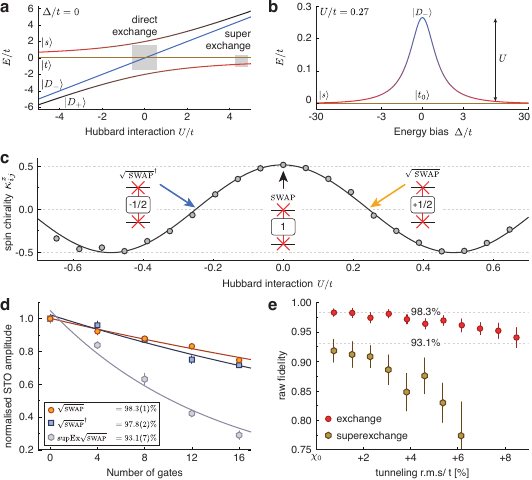}
         \caption{\textbf{Entangling gates via dynamical phases in the direct exchange regime}. (a) Energy spectrum of the two-particle Hilbert space as a function of the Hubbard energy $U$.
         (b) The spectrum for fixed $U/t=0.27$ as a function of the energy bias $\Delta/t$, for which we realise the \sswap gate. The acquired dynamical phase $\delta$ is equal to the integrated energy difference between the triplet $\ket{t_0}$ and the state $\ket{\psi}$ (red/blue line). (c) Calibration of the gate exponent $\alpha$ through measurement of the spin chirality $\kappa_{ij}^z$ as a function of the Hubbard interaction $U$. The data points are the amplitudes of sinusoidal fits to the STOs of $4\,\text{-}\,7$ experimental repetitions, while the error bars (fit uncertainties) are smaller than the data points (see Methods - \ref{subsec:STO_Fidelity_correction}). The zero crossing at negative (positive) Hubbard interaction $U$ denote the \swapd (\sswap) gate. (d) Raw gate fidelities of the \sswap (orange circles) and \swapd (blue squares) gate in comparison with a \sswap gate realised in the superexchange regime (grey hexagons). The gate fidelity is obtained by fitting an exponential decay function to the data obtained from repeated sinusodial fits of STOs of $I=9\,\text{-}\,10$ experimental realisations (see Methods - \ref{subsec:STO_Fidelity_correction}). The data points (errorbars) are determined by the amplitude value (uncertainty) given by the fitting routine. (e) Noise measurement. We compare the resilience against tunnelling r.m.s noise applied to the lattice potential of the exchange (red circles) and superexchange (brown hexagons) gates. The dotted lines indicate the unperturbed gate fidelity without any added tunnelling noise. We plot the raw fidelity for $10\,\text{-}\,11$ experimental realisations, with error bars showing the standard error.
        }
    \label{fig:4}
\end{figure}

So far, our considerations have been limited to the geometric \swap gate realised in the noninteracting regime ($U=0$), facilitating resilient transport and rearrangement of atoms in dense qubit systems.
We now extend this technique into the interacting regime ($U\neq0$) to realise a tunable two-qubit entangling gate based on interatomic collisional processes (cf.~refs.~\cite{jaksch_entanglement_1999,brennen_quantum_1999,sorensen_spin-spin_1999,calarco_quantum_2000,hayes_quantum_2007,daley_quantum_2008,weitenberg_quantum_2011,nemirovsky_fast_2021,zhou_scheme_2022,singh_optimizing_2025}).
Utilising doublon states allows faster gate operation in the direct exchange regime ($\lvert U\rvert \leq t$, Fig. \ref{fig:4}a, and ref.~\cite{burkard_semiconductor_2023}), where the relevant energy scale is a first-order effect directly proportional to the Hubbard $U$ (Fig. \ref{fig:4}b).
This approach differs fundamentally from previous implementations in the superexchange ($U\gg t$, refs.~\cite{chalopin_optical_2025,zhu_splitting_2025}) or combined ($U\gtrsim t$,~\cite{yang_cooling_2020,zhang_scalable_2023,bojovic_high-fidelity_2025}) regimes, where second-order processes $\propto t^2/U$ become important.
Superexchange gates are inherently sensitive to fluctuations in tunnelling and rely on the suppression of  doublon states, requiring fine-tuned parameters or slow ramps.
In contrast, the direct exchange mechanism embraces the dynamics within the dark space manifold, including doublons, as an integral part of the operation.
To provide intuition for the involved processes, we plot the energy spectrum of the Fermi-Hubbard dimer as a function of the Hubbard interaction $U$ in Figure \ref{fig:4}a.
The total phase accumulated during a gate cycle $\varphi=\delta+\gamma$ defines the state transformation of the state $\ket{\psi}\rightarrow e^{i\varphi}\ket{\psi}$.
The dynamical phase $\delta$ is given by the time-integral of the energy along the evolution path of the state $\ket{\psi}$ and can be calculated as $\delta=\int_{\tau} E_{\psi}(\tau')d\tau'$ for a given gate duration $\tau$.
The energy of the state $\ket{\psi}$ as a function of the energy bias $\Delta/t$ is shown in Figure \ref{fig:4}b for $U/t=0.27$.
The energy of the triplet states $\mathcal{T}$ (including $\ket{t_0}$) remains zero for all settings of $\Delta/t$ due to fermionic exchange statistics.
In consequence, the relative dynamical phase accumulated between the two states is solely determined by the energy of the state $E_{\psi}$ during the gate operation which is given by the Hubbard interaction $U$ and the interaction time $\tau$.
Since $U$ is precisely tuneable via a Feshbach resonance~\cite{chin_feshbach_2010}, the energy of $\ket{\psi}$ can be both positive ($U>0$) and negative ($U<0$).
The default operation of our scheme is the \swap gate transforming $\ket{i_-}\rightarrow\ket{i_+}$.
The combination of the two phases $\gamma+\delta$ and their tunability results in tunable partial \swapa gates.

To calibrate the \swapa gate operations, we measure the gate exponent $\alpha$ for different values of the Hubbard interaction $U$. Starting from the input state $\ket{i_-}$, we apply a single gate at a chosen value of $U$ and subsequently measure the spin chirality $\kappa^z_{ij}=\langle (\Vec{S_i}\times\Vec{S_j})^z\rangle=\langle S_i^xS_j^y-S_i^yS_j^x\rangle$.
The spin chirality directly yields the exponent $\alpha$ according to the relation $\alpha(\kappa_{ij}^z)=\pm\arccos{(-2\kappa_{ij}^z)}/\pi$ for $\kappa_{ij}^z\in [0,0.5]$. 
Explicitly, it takes the value of $\pm0.5$ for the $\ket{i_{\mp}}$ states and zero for all states in $\{\ket{\uparrow,\downarrow},\ket{\downarrow,\uparrow},\ket{s},\ket{t_0}\}$. This quantity is only well defined at the start and end of the gate operation. It is measured by sampling the coherent STO at its respective maximum and minimum (see Methods - \ref{subsec:STO}).
In Figure \ref{fig:4}c we plot the measured spin chirality $\kappa^z_{ij}$ as a function of the Hubbard interaction $U$. The zero crossings of the spin chirality $\kappa_{ij}^z$ correspond to the \sswap and \swapd gates in our experiment. For $U=0$, we recover the geometric \swap-gate ($\kappa_{ij}^z=0.5, \alpha=1$).
To characterise the performance of the \sswap ($\alpha=1/2)$ and \swapd ($\alpha=-1/2$) gates, we measure their fidelity by applying a variable number of gates and quantifying the fidelity of returning to the initial state by measuring the amplitude of the subsequent STO.
In Figure \ref{fig:4}c, we plot the nomalised STO amplitude as a function of the number of applied gates.
We determine the fidelity of the gate operations by extracting the decay constant from an exponential fit.
The measured raw fidelities are given in the inset of the figure. Accounting for the dispersive loss of atoms, we can calculate loss-corrected fidelities of $99.0(2)\%$ (\sswap) and $98.6(2)\%$ (\swapd) (see Methods -  \ref{subsec:STO_Fidelity_correction}).
We compare these two gates realised in the direct exchange regime with a \sswap realised with the conventional method using the superexchange interactions ($U/t \approx 4.3$) by ramping the tunnelling barrier height with a Blackman pulse~\cite{kasevich_laser_1992} to engineer the desired interaction energy.
We measure a loss-corrected fidelity of the superexchange gate of $93.8(7)\%$. We attribute the significantly improved gate fidelities in the direct exchange regime to the dynamical phase contribution being proportional to $J_{\text{Ex}}\sim U$, unlike the superexchange energy, which depends on both tunnelling and interaction ($J_{\text{SupEx}}\sim t^2/U$).
We also compare the noise resilience of the \sswap gates realised in the direct and superexchange regimes, respectively.
We apply white noise of a bandwidth of $\SI{2}{\kilo\hertz}$ and variable amplitude during the operation of eight gates to $V_X$.
In Figure \ref{fig:4}e, we plot the raw fidelity as a function of the added tunnelling  noise (see Methods - \ref{subsec:STO_Fidelity_correction}).
While the superexchange \sswap suffers from the additional tunnelling noise, the direct-exchange \sswap displays a plateau up to $\SI{3}{\percent}$ added noise.
This quantitative comparison between the two methods under otherwise identical conditions---same apparatus, same lattice depth, and same gate duration $\tau$---highlights the robustness of direct exchange gates against fluctuations in the tunnel coupling.

\section{Conclusion and Outlook}

In summary, we demonstrate a high-fidelity geometric \swap gate by leveraging the physics of doublon states.
Here, two qubits occupy the same site in a lattice, naturally extending the computational Hilbert space.
This allows us to trace a holonomic loop that performs the \swap operation, whose fidelity is protected by global geometric properties of the path rather than the precise, fine-tuned dynamics of its execution.
In our experiment using fermionic atoms in a densely-spaced optical lattice, we achieve a loss-corrected amplitude fidelity of $99.91(7)\%$.
The residual \swap error is caused mainly by fluctuating Hubbard $U$ (Methods - \ref{subsec:noise_contributions}), which can further improved by magnetic field stabilisation or by using an atom whose scattering length is less sensitive to changes of the magnetic field, suggesting avenues to even higher fidelity.

\swap gates are of central importance in quantum codes~\cite{fernandez_implementing_2024,andersen_small_2025} and algorithms~\cite{tan_optimal_2020,camps_quantum_2021,blekos_review_2024,tan_optimal_2020}, since they enable nonlocal operations and quantum information routing in densely-packed qubit arrays.
A competing approach towards nonlocal connectivity is moving qubits, for instance using atoms in steerable tweezers~\cite{evered_high-fidelity_2023,rines_demonstration_2025}, but this leads to overheads due to the necessary empty space between the qubits.
Conversely, direct \swap implementations are still challenging for quantum computing architectures which do not intrinsically support exchange-type physics.
For instance, realising qubit \swap using Rydberg gates would require concatenating three CZ and six Hadamard gates.
Taking into account two-qubit gate fidelities of around $99.6\%$~\cite{evered_high-fidelity_2023,peper_spectroscopy_2025,muniz_high-fidelity_2025,senoo_high-fidelity_2025,rines_demonstration_2025} the composite \swap gate fidelity would drop below $99\%$, further highlighting the advantage of \swap gates built from qubit doublons.

We also demonstrate entangling \swapa gates which are faster and more robust than the state-of-the-art superexchange gates thanks to the population of doublon states. 
The doublon approach is also suited for fermionic quantum processing~\cite{bravyi_fermionic_2002,gonzalez-cuadra_fermionic_2023}, as it operates in the motionally coherent regime.
It integrates with topological pumping for nonlocal connectivity~\cite{zhu_quantum_2025} and enhances the feasibility of crucial quantum algorithms in densely-spaced qubit arrays~\cite{tan_optimal_2020,camps_quantum_2021,blekos_review_2024,tan_optimal_2020,fernandez_implementing_2024,andersen_small_2025}.
The underlying physics of the geometric gate is platform-independent, suggesting its implementation in other quantum architectures, such as semiconductor spin qubits~\cite{burkard_semiconductor_2023,kaplan_spin_2004,hu_gate_2002,zhou_high-fidelity_2025} and Rydberg atom arrays~\cite{evered_high-fidelity_2023,peper_spectroscopy_2025,muniz_high-fidelity_2025,senoo_high-fidelity_2025,rines_demonstration_2025}, to alleviate space constraints in an all-to-all coupled array.
More broadly, our work highlights the role of global system properties for quantum control, such as Hamiltonian symmetry, quantum geometry, and exchange statistics, in the same spirit as the robustness of topological matter~\cite{chiu_classification_2016}.\\

During preparation of the manuscript, we became aware of a conceptually related work on two-particle holonomies with photon waveguides~\cite{neef_pairing_2026}.\\

\section*{Acknowledgements}

We thank Alexander Frank for assistance with electronics equipment.
We acknowledge funding by the Swiss National Science Foundation (Grant No.~200020\_212168, Advanced grant TMAG-2\_209376, 20QT-1\_205584, as well as Holograph UeM019-5.1).
Y.K.~acknowledges funding from the ETH Postdoctoral Fellowship 24-2 FEL-035.


%

\clearpage

\setcounter{figure}{0} 
\setcounter{equation}{0}

\renewcommand{\figurename}{Extended Data Fig.}
\renewcommand{\tablename}{Extended Data Table}
\renewcommand\thefigure{\arabic{figure}} 
\renewcommand\thetable{\arabic{table}} 
\renewcommand\theequation{M\arabic{equation}}

\section*{Methods}

\renewcommand{\thesection}{}

\subsection{Dynamical superlattice potential}
\label{subsec:SLPotential}
The optical lattice in our setup is generated by a single red-detuned laser with $\lambda=1064\,\text{nm}$, retro-reflected in all three spatial dimensions. Along the $x$-direction, a second beam ($X_\mathrm{int}$) is superimposed to create an interference pattern with the beam in the $z$-direction. The relative phase of the $X_\mathrm{int}$ and $Z$ lattice beams $\varphi_\text{SL}(\tau)$ can be adjusted dynamically, resulting in the time-dependent potential given by   
\begin{equation}
\begin{split}\label{eq:potential}
    V(&x,y,z,\tau) = \\
    &-V_\mathrm{X}\cos^2(kx+\theta/2)\\
    &-V_\mathrm{Xint}\cos^2(kx)\\
    &-V_\mathrm{Y}\cos^2(ky)\\
    &-V_\mathrm{Z}\cos^2(kz)\\
    &-\sqrt{V_\mathrm{Xint}V_\mathrm{Z}}\cos(kz)\cos(kx+\varphi_{\text{SL}}(\tau))\\
    &-I_\mathrm{XZ}\sqrt{V_\mathrm{Xint}V_\mathrm{Z}}\cos(kz)\cos(kx-\varphi_{\text{SL}}(\tau)),
\end{split}
\end{equation}
where $k=2\pi/\lambda$, the imbalance factor is $I_\text{XZ}$, and $\{V_\mathrm{X},V_{\mathrm{X}_\text{int}},V_\mathrm{Y},V_\mathrm{Z}\}$ denote the lattice depth of the individual beams with the values used in the experiment listed in Extended Data Table \ref{tab:lattice_depths}. The tunnelling dynamics along the $y$- and $z$-directions are frozen out on relevant timescales by choosing a sufficiently deep lattice potential in these spatial directions, effectively creating independent one-dimensional arrays along the $x$-direction. Each one-dimensional system is characterised by the tunable bias $\Delta$ and tunnellings $t$ and $t'$ (see Extended Data Fig.~\ref{fig:Super-Exchange_lattice_parameters}a). 
Cyclic modulation of $t$($t'$) and $\Delta$ is implemented by linearly ramping the superlattice phase $\varphi_\text{SL}(\tau)$ with an acousto-optic modulator acting on the $V_\mathrm{Xint}$ beam~\cite{walter_quantization_2023}.

{\renewcommand{\arraystretch}{1.2}
\begin{table*}[ht]
    \centering
    {\fontfamily{phv}\selectfont 
    \begin{tabular}{lcllll}
        \hline\hline
        & $V_\mathrm{X}$[$E_\mathrm{rec}$] & $V_\mathrm{Xint}$[$E_\mathrm{rec}$] & $V_\mathrm{Y}$[$E_\mathrm{rec}$] & $V_\mathrm{Z}$[$E_\mathrm{rec}$] & $I_\mathrm{XZ}$\\
        \hline
        Fig.~\ref{fig:2}(a,b) $\ket{\psi(\Delta/t)}$ & 10.10(6)$^{\text{\tiny{a}}}$ & 1.00(3) & 31.10(6) & 29.54(2) & 0.804(1)\\
        Fig.~\ref{fig:2}(d) f-\swap STO & 9.93(7) & 0.99(3) & 31.15(8) & 30.10(2) & 0.800(1)\\
        Fig.~\ref{fig:3}(a) geom. \swap & 9.97(8) & 1.02(3) & 31.07(6) & 30.27(9) & 0.800(1)\\
        Fig.~\ref{fig:3}(b) geom. \swap noise & 9.99(6) & 0.97(2) & 31.23(6) & 21.56(7) & 0.758(1)\\
        Fig.~\ref{fig:4}(c) $\alpha$-calibration & 9.96(8) & 1.02(3) & 31.09(5) & 30.24(11) & 0.800(1) \\
        Fig.~\ref{fig:4}(d) \sswap & 10.02(8) & 1.06(3) & 31.04(5) & 29.92(5) & 0.800(1) \\
        Fig.~\ref{fig:4}(d) \swapd & 9.97(7) & 1.00(2) & 31.08(6) & 30.03(8) & 0.800(1) \\
        Fig.~\ref{fig:4}(d) SupEx\sswap & 24.9(1)/9.89(6)$^{\text{\tiny{b}}}$ & 0.97(3) & 31.15(4) & 21.70(10) & 0.758(1) \\
        Fig.~\ref{fig:4}(e) noise Ex & 10.09(7) & 1.03(2) & 31.23(7) & 21.47(7) & 0.758(1) \\
        Fig.~\ref{fig:4}(e) noise SupEx & 25.1(1)/10.07(6)$^{\text{\tiny{b}}}$ & 1.01(2) & 31.05(5) & 21.50(5) & 0.758(1) \\
        \hline
    \end{tabular}
    }
    \caption{\textbf{Lattice depths used in the experiment.} $^{\text{\tiny{a}}}$The errors in the brackets correspond to the statistical standard error. $^{\text{\tiny{b}}}$The two values of $V_{\text{X}}$ for superexchange (SupEx) gate correspond to the maximum and minimum values of the Blackman pulses.}
    \label{tab:lattice_depths}
\end{table*}}

\subsection{Singlet-triplet oscillations}
\label{subsec:STO}

\begin{figure}[htbp]
    \centering
    \includegraphics[width=\linewidth]{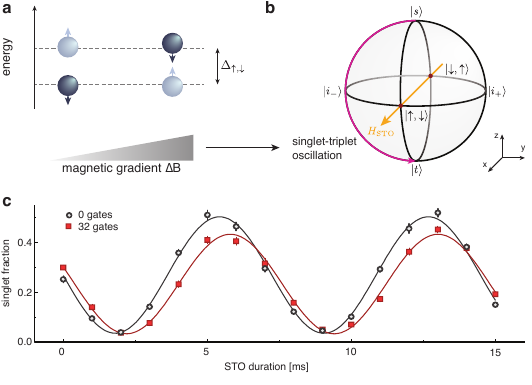}
    \caption{\textbf{Singlet-triplet-oscillation (STO) measurement protocol and two-particle Bloch sphere.} (a) Applying a magnetic field gradient along the $x$-direction induces an energy shift between $\ket{\uparrow,\downarrow}$ and $\ket{\downarrow,\uparrow}$. (b) The energy offset dynamically couples the singlet and triplet states visualised as a rotation on the two-particle Bloch sphere around the $x$-axis. (c) The time trace of the singlet fraction with an applied magnetic field gradient. The amplitude of the STO signal determines the returning fidelity after $N$ gate operations, starting from the $\ket{i-}$ state. The decay of the oscillation offset is a result of the dissipative loss of atoms in the lattice. A phase shift is induced by a light shift in the optical lattice, which can be corrected for using standard dynamical decoupling protocols. }
    \label{fig:STO_measurement_decay}
\end{figure}

A central tool for state preparation, readout, and coherent manipulation in our experiment are singlet-triplet oscillations (STOs) of two-particle states. Applying a magnetic field gradient along the x-direction (Figure~\ref{fig:STO_measurement_decay}a) lifts the degeneracy between the $\ket{\uparrow,\downarrow}$ and $\ket{\downarrow,\uparrow}$ states, resulting in coherent population transfer between the singlet $\ket{s}$ and triplet $\ket{t_0}$ states. In a reduced two-particle Hilbert space, this can be visualised as a rotation on the two-particle Bloch sphere as shown in Extended Data Fig.~\ref{fig:STO_measurement_decay}b. The corresponding dynamics is governed by an effective two-level Hamiltonian in the reduced Hilbert space spanned by $\{\ket{s},\ket{t_0}\}$ which can be written as
\begin{equation}\label{eq:STO_hamiltonian}
    \hat{H}_\text{STO} = \frac{1}{2} \begin{pmatrix}
        0 & \Delta_{\uparrow\downarrow} \\ \Delta_{\uparrow\downarrow} & 0 
    \end{pmatrix}. 
\end{equation}
In our experiment, the application of a gradient of $\Delta B \simeq \SI{3.8}{\gauss\per\centi\meter}$ induces coherent STOs at a frequency of $f_\text{STO} = \Delta_{\uparrow\downarrow}/h \simeq 140\,\text{Hz}$ for singlets separated by one lattice site. Therefore, the half-oscillation time $\tau_{\text{STO}}/2 \approx  3.5\,\text{ms}$ defines the STO duration required to rotate the singlet state to the triplet state.
The amplitude and phase of the oscillation are used to infer the two-particle state on the Bloch sphere (Extended Data Fig.~\ref{fig:STO_measurement_decay}c). A sinusoidal fit of the form 
\begin{equation}\label{eq:STO_fit}
    f(T)=A_\mathrm{STO}\sin(f_\mathrm{STO}\,T+\varphi)+y_0
\end{equation}
extracts the STO amplitude $A_\mathrm{STO}$, the oscillation offset $y_0$ and phase $\varphi$, which are used for calibration and fidelity measurements (see following sections). 

\subsection{State preparation}
\label{subsec:statePrep}

\begin{figure}[htbp]
    \includegraphics[width=0.48\textwidth]{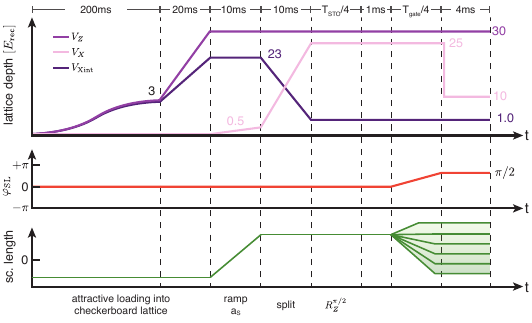}
         \caption{\textbf{Experimental Sequence for $\ket{i-}$ state preparation.} Loading procedure to prepare atom pairs in $\ket{i-}$ states in the dynamical superlattice. 
         The atoms are loaded into the unit cells of a shallow checkerboard lattice under strongly attractive interactions ($a_{\textbf{S}}<0)$. The lattice potentials $V_\text{Z}$ and $V_\text{Xint}$ are subsequently increased, after which the scattering length is tuned to a positive value, followed by the splitting of each single well into a double well. $V_\text{Y}$ is first ramped to 7 $E_r$ in 200 ms and then to 31 $E_r$ in 20 ms, similar to $V_\text{Z}$ and not shown in the figure. At last, the singlet pairs are transferred into $\ket{i-}$ with a quarter STO.}
         \label{fig:S1_StatePrep}
\end{figure}

We prepare an evaporatively cooled spin mixture of $5.8(2)\times10^4$ potassium-40 atoms ($F = 9/2$, $m_F = \{-9/2, -7/2\}$) at a temperature of $0.089(3)\,T_F$ in a crossed optical dipole trap.
Atomic singlet pairs are then loaded into the optical lattice following the experimental sequence shown in Extended Data Fig.~\ref{fig:S1_StatePrep}. The loading starts with strongly attractive interactions using the Feshbach resonance between internal states $m_F = -9/2$ and $-7/2$ located at $\SI{201.1}{\gauss}$. A shallow checkerboard lattice is first ramped to $\sim3\, E_r$ over $\SI{200}{\milli\second}$, where $E_r=h^2/(2m\lambda^2)$ is the lattice recoil energy. This is followed by a more rapid increase to a deep lattice potential within $\SI{20}{\milli\second}$, after which the scattering length is tuned to large positive values, corresponding to repulsive interactions. Then, gradually increasing $V_\mathrm{X}$ and reducing $V_\mathrm{Xint}$, the single well of the checkerboard lattice is split into two sites of the double well, with one atom occupying each site.
The singlet state is characterised by spin correlations between the atoms in the left and right sites of the double well.
Following this protocol, we prepare about $60\,\text{-}\,70$\% of the atoms in the singlet state $\ket{s}$, while the remaining atoms are unpaired, occupying singly filled double wells (singlons). While this fraction depends on the initial atom number and temperature, it is independent of the lattice parameters used during the gate operation. We note that finite temperature and entropy only manifest in the filling, not in the singlet fidelity of the paired atoms in the same dimer. This is because our preparation scheme is based on splitting doublons in the checkerboard lattice, which are guaranteed to be in the spin singlet state due to fermionic anti-symmetrization. The measured gate fidelity is independent of the initial preparation efficiency because the gate protocol does not measurably affect the remaining unpaired atoms (singlons). In view to realising a fully programmable quantum computer we emphasise that our state preparation scheme should be further improved. 
Recent work using potential shaping techniques in a similar lattice configuration has demonstrated preparation fidelities exceeding 99\%~\cite{xu_neutral-atom_2025}.

For fidelity measurements we bring the two-particle state $\ket{s}$ to the $|i_-\rangle = (\ket{t_0}-i\ket{s})/\sqrt{2} = (\ket{\downarrow,\uparrow}-i\ket{ \uparrow,\downarrow})/\sqrt{2}$ state, by inducing a quarter-STO with the magnetic gradient for a time $\tau_\text{STO}/4\approx\SI{1.8}{\milli\second}$.
Subsequently, we ramp the magnetic field to a fixed value which corresponds to a specific gate operation during the experimental sequence. Before the gate sequence, the superlattice phase $\varphi_\text{SL}$ is adjusted to $\pi/2$ and $V_\mathrm{X}$ is quenched to $10 \,E_r$ to create the biased double well potential. We then allow the magnetic fields to stabilise with a hold time of $4\, \text{ms}$.

\subsection{Experimental sequence}
\label{subsec:expSeq_ReadOut}
\begin{figure}[tbp]
    \includegraphics[width=0.48\textwidth]{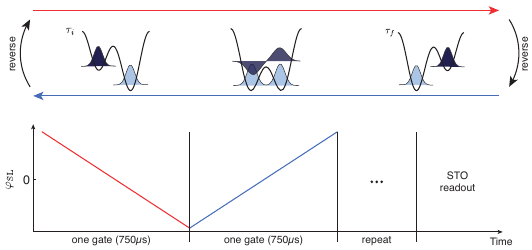}
         \caption{\textbf{Experimental sequence for consecutive gate operations.} Ramping the superlattice phase $\varphi_\text{SL}$ back and forth between $\pi/2$ and $-\pi/2$ realises consecutive gate operations on the same atomic pair.}
    \label{fig:S4}
\end{figure}

We discuss the experimental protocol used to measure the gate fidelity presented in Figures \ref{fig:3}a, \ref{fig:3}b and \ref{fig:4}d,e. 
We prepare atomic pairs in the $\ket{i-}$ state in the double wells and adjust the interaction $U$ to implement the desired gate. A single gate operation is performed by applying a phase ramp of $\varphi_\text{SL}$ from $\pm\pi/2$ to $\mp\pi/2$.
It does not require any ramp of lattice laser intensity.
To realise circuits of multiple two-qubit gates on the same pair, we reverse the phase ramp after one gate and return to the initial configuration of the potential realizing a total of two gates. This process is repeated until the desired number of gates is applied (see Extended Data Fig.~\ref{fig:S4}). 

To prevent unwanted additional dynamics after the gate sequence, we stop the phase ramp in the staggered configuration of the lattice, immediately followed by quenching into a deep cubic lattice ($V_\mathrm{X,Y,Z}>30\,E_r$, $V_\mathrm{Xint}=0\,E_r$) within $\SI{100}{\micro\second}$ to freeze the tunnelling dynamics. 
Before STO detection, we remove atoms on doubly-occupied lattice sites with two consecutive Landau-Zener RF-sweeps, transferring atoms in the $m_F=-7/2$ to $-3/2$. This activates spin-changing collisions with the $m_F=-9/2$ atoms and causes pairs to leave the trap. The remaining $-3/2$ population is then transferred back to $-7/2$. Next, the STO sequence described in Section~\ref{subsec:STO} is performed, followed by merging the double wells into single wells, where the singlet is converted into double occupancy in the ground state and subsequently measured by orbital-selective RF spectroscopy~\cite{walter_quantization_2023}. 

\subsection{Gate fidelity measurement and correction}
\label{subsec:STO_Fidelity_correction}

To determine the fidelity of the gate mechanism, we measure singlet-triplet oscillations (STO) (Section~\ref{subsec:STO}) after a variable number of consecutive $N$ gates. From a sinusoidal fit of the form 
\begin{equation}\label{eq:STO_fit_simpl}
    f(T) = A\sin{(f\,T+\varphi)} +y_0
\end{equation}
we extract the amplitude $A$ and the offset $y_0$ of the STO. All uncertainties obtained from the fitting procedures in this work are based on a weighted least-squares method, where each data point is weighted by $1/\sigma_i^2$, with $\sigma_i$ denoting the standard error of the $i$-th data point.
Gate errors in the exponent $\alpha$ manifest as deviations of returning to the initial state $\ket{i_-}$ on the equatorial plane of the $\ket{s}$ - $\ket{t_0}$ Bloch sphere and therefore lead to a decrease of the STO amplitude $A$. 

By fitting an exponential decay $f(N)=A_1\exp{(-N/N_e)}$ to the amplitude as a function of number of applied gates, we obtain a decay constant $N_e$. The raw fidelity of the gate is then calculated as $\mathcal{F}_\mathrm{raw}=\exp{(-1/N_e)}$, with errors propagated from the exponential fitting procedure.

Due to finite residual tunnelling to neighboring double wells (dispersive atom loss), we observe a decay of the the offset $y_0$ for an increasing number of gates, while the minimum of the STO remains constant. 
To account for this effect, we analyze the offset $y_0$ as a function of the applied gate number using the same exponential fit to extract a second decay constant $O_e$ that characterizes the survival fidelity $\mathcal{F}_\mathrm{surv} = \exp{(-1/ O_e)}$ (e.g., Figure \ref{fig:3}a). To calculate the loss-corrected fidelity, we divide the raw fidelity by the survival fidelity $\mathcal{F}_\mathrm{corr} = \mathcal{F}_\mathrm{raw}/\mathcal{F}_\mathrm{surv}$. The loss corrected amplitude shown in Figure~\ref{fig:3}a is calculated with the STO amplitude $A$ divided by the offset $y_0$, and the corrected fidelity $\mathcal{F}_\mathrm{corr}$ is obtained directly from the fit of the corrected amplitude. This yields the same fidelity value as the method described above. The fit errors are obtained from the fitting procedure.

When probing the robustness of the \swap gate in Figure \ref{fig:3}b, the STO amplitude $A$ is obtained from a two-point measurement of the STO minimum and maximum. Each data point and the corresponding error bar in Figure \ref{fig:3}b is calculated with $\mathcal{F}_{\text{raw}}=(A_{N}/A_{0})^{1/N}$, where $A_{N}$ is the two-point amplitude after $N=16$ gates. For the dynamical gates investigated in Figure \ref{fig:4}e, we use the same procedure but apply the noise only during a total of $N=8$ gates. 

\subsection{Geometric phase}
\label{subsec:Geometric_phase}

During gate operation, the singlet state $\ket{s}$ acquires a phase relative to the fully decoupled triplet states $\mathcal{T}$. The phase can be purely geometric (Aharonov-Anandan phase \cite{aharonov_phase_1987}) or have additional dynamical phase contributions with finite $U$. 
If the evolution of the Hamiltonian $H(\tau)$ satisfies the adiabatic theorem, a system prepared in an initial eigenstate $\ket{\varphi(\tau_i)}$ will remain in the corresponding instantaneous eigenstate throughout the process. 
Adiabatic quantum gates are defined by the overall transformation of the quantum state from the initial superposition $\ket{\Psi(\tau_i}=\sum_n a_n\ket{\varphi_n(\tau_i)}$ to the final state $\ket{\varphi(\tau_f)}=\sum_n a_n e^{i\Gamma_n}\ket{\varphi_n(\tau_f)}$, where each eigenstate component acquires a phase $\Gamma_n$ along its adiabatic path. For an evolution, where the eigenbasis returns to its initial configuration, i.e. $\ket{\varphi_n(\tau_f)}=\ket{\varphi_n(\tau_i)}$, the gate is fully characterised by the set of phases $\{\Gamma_n\}$ giving the unitary transformation performed by the evolution. The total acquired phase per basis state $\Gamma_n$ is given by the sum of the well-known dynamical phase $\delta$ 
\begin{equation}\label{eq:dynamical_phase}
    \delta = \int_{\tau_i}^{\tau_f} E_n(\tau')d\tau',
\end{equation}
where $E_n(\tau)$ is the eigenenergy of state $\ket{\varphi_n(\tau)}$ and the geometric phase defined as \cite{aharonov_phase_1987} 
\begin{equation}\label{eq:geometric_phase}
    \gamma = \int_{\tau_i}^{\tau_f} \bra{\varphi_n(\tau')} \frac{d}{d\tau'}\ket{\varphi_n(\tau')} d\tau'.
\end{equation}
Note that while the geometric phase is gauge independent, the above expression is valid only for a closed trajectory in the state space and a gauge that satisfies $\ket{\varphi_n(\tau_f)}=\ket{\varphi_n(\tau_i)}$. 
To evaluate the geometric phase using Eq.~\ref{eq:geometric_phase}, we compute the instantaneous eigenstate
\begin{equation}\label{eq:eigenstate}
\ket{\varphi_{\text{gauge}}(\tau)}=e^{i\beta(\tau)}\left[\cos \left(\frac{\theta(\tau)}{2}\right)\ket{s}+\sin \left(\frac{\theta(\tau)}{2}\right)\ket{D_-}\right].
\end{equation}
For the initial and final configurations we have $\theta(\tau_i)\simeq0$ and $\theta(\tau_f)\simeq 2\pi$. 
The prefactor $e^{i\beta(\tau)}$ serves solely to validate Eq.~\ref{eq:geometric_phase} by enforcing the gauge condition $\ket{\varphi_{\text{gauge}}(\tau_f)} = \ket{\varphi_{\text{gauge}}(\tau_i)}$.
By substituting Eq.~\ref{eq:eigenstate} into Eq.~\ref{eq:geometric_phase}, we obtain \( \gamma = \beta(\tau_i) - \beta(\tau_f) \), which is quantised to $\pi$ modulo $2\pi$ due to the gauge constraint.
If dynamical phases are absent, this realises a purely geometric \swap gate in the computational space $\mathcal{C}$ represented by the matrix 
\begin{equation}\label{eq:swap_matrix}
    \hat{U}_{\swap} = \begin{pmatrix}
        1&0&0&0\\
        0&0&1&0\\
        0&1&0&0\\
        0&0&0&1
    \end{pmatrix}\,.
\end{equation}

\subsection{Symmetries of the Hamiltonian}
\label{subsec:ExGate_robust}

The Hamiltonian of our system has two important symmetries: chiral and time-reversal symmetry. Both ensure protected state evolution, while the optical potential is dynamically modulated. In the basis of spin operator eigenstates spanned by the states $\{\mathcal{T}, \mathcal{S}\}$, the Hamiltonian takes the form 
\begin{align}\label{eq:Hsinglet_Methods}
    \hat{H}_{\mathcal{T},\mathcal{S}}(\tau)=\begin{pmatrix}
        \mathbf{0_3}& &\mathbf{0_3} & \\
        &U & 2\Delta(\tau) & -2t(\tau) \\
        \mathbf{0_3}& 2\Delta(\tau) & U & 0 \\
        & -2t(\tau) & 0 & 0 
        \end{pmatrix}, 
\end{align}
with two distinct cases: $U\neq0$ and $U=0$. For the non-interacting scenario, corresponding to the purely geometric \swap gate, the Hamiltonian possesses both chiral and time-reversal symmetry. For non-vanishing Hubbard interactions $U$, only the time-reversal symmetry is fulfilled. 

The real and symmetric Hamiltonian is time-reversal symmetric with respect to the anti-unitary operator $T=-(\mathbf{I_3}\otimes\sigma_z)K$, where $K$ denotes complex conjugation~\cite{chiu_classification_2016}. This symmetry ensures that all eigenstates can be chosen to be real throughout the evolution. Therefore, a system prepared in an eigenstate remains within the real-valued subspace, provided that the evolution is adiabatic. This also ensures that the relative phase between $\ket{s}$ and $\ket{D_-}$ in the dark state $\ket{\psi}$ is quantized to $0$ or $\pi$, and thus the trajectory of the state is constrained to the meridian of the singlet Hilbert space in Figure \ref{fig:1}d. 
During the gate operations, the time-reversal symmetry ensures that the adiabatic path remains confined to this subspace. As a result, noise or imperfection in the control Hamiltonian that respect time-reversal symmetry cannot cause leakage from this protected subspace. This constraint implies that the geometric phase acquired over a closed cycle in Hilbert space is quantised and real ($0$ or $\pi$). The trajectory in Hilbert space is thus symmetry protected.

In the non-interacting case ($U=0$), the Hamiltonian in equation \eqref{eq:Hsinglet_Methods} has an additional chiral symmetry with respect to the unitary operator $\Gamma=\mathrm{diag}(\pm1,\pm1,\pm1,+1,-1,-1)$, satisfying the anti-commutation relation $\Gamma\hat{H}_{\mathcal{T},\mathcal{S}}\Gamma^{-1}=-\hat{H}_{\mathcal{T},\mathcal{S}}$. 
The $\mathcal{T}$ states form each an isolated one-dimensional subspace that commutes trivially with both $\hat{H}_{\mathcal{T},\mathcal{S}}$ and $\Gamma$, preserving their zero-energy eigenvalues. The remaining three-dimensional $\mathcal{S}$ subspace is constrained to have eigenvalues that appear in symmetric pairs with $\pm E$. Since this subspace has odd dimensionality, one eigenvalue must remain at $E = 0$~\cite{chiu_classification_2016}.
In the analysed system, the zero-energy eigenvalue belongs to the dark state $\ket{\psi}$, ensuring the pure-geometric nature of the \swap gate. As the symmetry is independent of the Hamiltonian parameters, it is also robust against fluctuations of control parameters $\Delta$ and $t$. 
As a consequence of the chiral symmetry, the Hilbert space is divided into two separate subspaces $A=\{\ket{D_+}\}$ and $B=\{\ket{D_-}, \ket{s}\}$. This separation can be understood directly from the structure of the Hamiltonian, which contains only off-diagonal couplings between the two subspaces. States with the same eigenvalue under $\Gamma$ are not directly coupled by the Hamiltonian. As a result, the Hamiltonian only allows transitions between states of opposite chiral symmetry, and the system's dynamics are confined to coherent processes between the two subspaces. The $\ket{s}$ and $\ket{D_-}$ states in the dark state $\ket{\psi}$ are thus indirectly coupled via $\ket{D_+}$, which remains always unpopulated due to destructive interference.

\subsection{Dynamical gate parameters}
\label{subsec:DynamicalGateLatticeParams}

\begin{figure}[tbp]
    \includegraphics[width=0.48\textwidth]{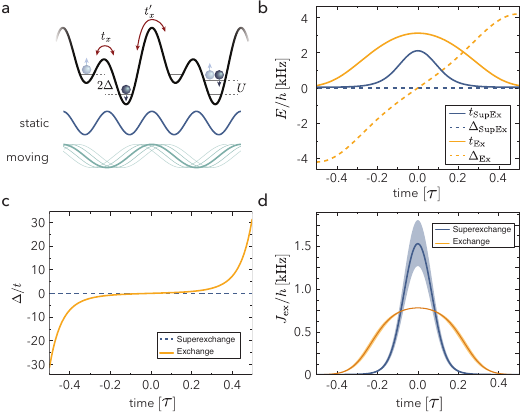}
         \caption{\textbf{Dynamical lattice potential and corresponding scaling of Hamiltonian parameters.} (a) A schematic of the parameters in the dynamical optical superlattice. (b) shows the time dependent tunnelling $t$ and bias $\Delta$ during the Hamiltonian evolution in the exchange and superexchange regimes for their respective gate implementations. In (c) the experimental sweep of the ratio $\Delta/t$ is shown, used for the $x$-axis in Figures \ref{fig:2}a, b and \ref{fig:4}b. 
         (d) are the (super-)exchange energies for the two gate mechanisms for the respective \sswap gates. The shaded background shows the fluctuation in the exchange energy if the tunnelling is changed by $\pm$10\%. The lattice parameters used for the calculation of superexchange interaction are $[V_{\text{X}}, V_{X_\text{int}}, V_{\text{Y}}, V_{\text{Z}}] = [9.9, 1.0, 31.1, 21.7]E_r$ with $I_{\text{XZ}}=0.76$ and $U_\text{SupEx}/h=9.6\,\text{kHz}$. In the exchange regime the experimental parameters are $[V_{\text{X}}, V_{X_\text{int}}, V_{\text{Y}}, V_{\text{Z}}] = [10.0, 1.1, 31.0, 29.9]E_r$, with $I_{\text{XZ}}=0.8$ and $U_\text{Ex}/h = 780\,\text{Hz}$.}
    \label{fig:Super-Exchange_lattice_parameters}
\end{figure}

By combining the geometric and dynamical gates the complete set of \swapa two-qubit gates is accessible. The matrix representation of this gate set in the computational basis $\mathcal{C}$ is given by 
\begin{align}
    \hat{U}_{\alpha} = \begin{pmatrix}
        1&0&0&0\\
        0&(1+e^{i\pi\alpha})/2 &(1-e^{i\pi\alpha})/2 &0\\
        0&(1-e^{i\pi\alpha})/2 &(1+e^{i\pi\alpha})/2 &0\\
        0&0&0&1
    \end{pmatrix},
\end{align}
where the gate is controlled by tuning $\alpha$ with varying dynamic $\delta$ and geometric $\gamma$ phase contributions (Methods \ref{subsec:Geometric_phase}). 

We directly compare two-qubit gates realised in the exchange and superexchange regimes. These regimes differ in their underlying physical properties and therefore require different experimental implementations. In the exchange regime (Section \ref{subsec:ExGate_robust}), we employ a protocol in which the bias $\Delta(\tau)$ and tunnelling $t(\tau)$ are dynamically modulated and the direct exchange energy is at first-order not proportional to $t$. In contrast, gate performance in the superexchange regime is more sensitive to fluctuations in the tunnelling amplitude. For the superexchange gate, we only modulate the barrier height, and thus the tunnelling $t$, between adjacent sites. The gate mechanism relies on rapid tuning of the tunnelling rate using a Blackman-shaped pulse for $V_{\text{X}}$ \cite{kasevich_laser_1992}. 

For the superexchange gates, the system is described by the Hamiltonian in equation \eqref{eq:Hsinglet_Methods}, with the bias $\Delta(\tau)$ fixed to zero. The relevant eigenstates that span this subspace are $\{\ket{s},\ket{t_0},\ket{t_+},\ket{t_-}\}$, with the triplet states $\ket{t_+}=\ket{\uparrow,\uparrow}$ and $\ket{t_-}=\ket{\downarrow,\downarrow}$. In contrast to the exchange regime, the protective chiral symmetry is always broken at large $U$. However, the ground state that accumulates the dynamical phase during the gate operation is still well approximated by the $\ket{s}$ state, as the contribution from $\ket{D_+}$ is strongly suppressed under large repulsive interactions.

To compare the gate robustness in the superexchange and direct exchange regimes, we consider dynamical phase-acquiring gates in both cases. To experimentally benchmark the robustness of the gates, we induce laser intensity fluctuations by superimposing additional white noise to the lattice laser beam intensity $V_X$ during the application of 8 gates. The additional noise has a bandwidth of $2\,\text{kHz}$ and is applied with varying amplitude (Figure \ref{fig:4}e). 

The enhanced robustness of gates in the direct exchange regime can be attributed to two factors. First, the direct exchange energy remains insensitive to first-order tunnelling effects from laser intensity noise, as its direct scaling only depends on $U$. In Extended Data Fig.~\ref{fig:Super-Exchange_lattice_parameters}d, we compare lattice imperfections in the direct exchange and superexchange gates for tunnelling fluctuations of $\pm10\%$. Both curves were obtained using the gate protocols previously described. The superexchange energy variations are amplified due to the proportionality $J_\mathrm{SupEx} \propto t^2$. In contrast, lattice laser power fluctuations have a minimal effect on the $J_\mathrm{Ex}$ energy acquired during direct exchange gate cycles.

A diagonalisation of the Hamiltonian in equation \eqref{eq:Hsinglet_Methods} for a vanishing Hubbard interaction ($U = 0$) yields several eigenvalues. In addition to the three zero-energy triplet states $\mathcal{T}$, we find non-zero eigenvalues of $E_{B',B}=\pm2\sqrt{\Delta^2+t^2}$ and a zero-energy eigenvalue corresponding to the dark state $\ket{\psi}$.
The non-zero energy eigenstates $\ket{\psi_{B,B'}}$ are thus separated from the zero-energy eigenstates by a gap $\Delta E\geq 2t$. 

For finite Hubbard interactions ($U\neq0$), a perturbative calculation of the eigenenergy of the dark state $\ket{\psi}$ yields the exchange energy in the regimes with strong and weak on-site interactions 
\begin{align}
    &E_{\ket{\psi}} \quad\overset{U\ll t,\Delta}{\longrightarrow} &&J_\mathrm{Ex}= \frac{U}{(\Delta/t)^2-1} \, ,\label{eq:Jexchange}\\
    &E_{\ket{\psi}} \quad\overset{U\gg t,\Delta}{\longrightarrow} &&J_\mathrm{SupEx} = \frac{4t^2}{U(1-(2\Delta/U)^2)} \,.\label{eq:Jsuperexchange}
\end{align} 
At $\Delta=0$ the (super)exchange interaction reach their maximum and the perturbative calculation recovers the well-known scaling of the superexchange energy $J_\mathrm{SupEx}=4t^2/U$, while for exchange interaction it is $J_\mathrm{Ex}=-U$, which is independent of $t$. This dependence strongly reduces the sensitivity of the direct exchange gate to tunnelling noise as seen in Extended Data Fig.~\ref{fig:Super-Exchange_lattice_parameters}d. 
While equations \eqref{eq:Jexchange} and \eqref{eq:Jsuperexchange} give a parametric scaling of the exchange energies $J_\text{(Sup)Ex}$, in Extended Data Fig.~\ref{fig:Super-Exchange_lattice_parameters} we numerically diagonalize the Hamiltonian in equation \eqref{eq:Hsinglet_Methods} to get the exact instantaneous eigenenergies.

\subsection{Noise contributions}
\label{subsec:noise_contributions}

The main contributions to gate noise come from laser intensity noise in the lattice beams, current fluctuations in the coils for the magnetic field, and inhomogeneities of the trapping potential. Fluctuations in the lattice laser intensity and inhomogeneity directly translate to tunnelling noise, which is theoretically analysed in Extended Data Fig.~\ref{fig:Super-Exchange_lattice_parameters}d for both gate implementations of \sswap in the tight-binding limit. Varying the tunnelling by $\pm10\%$ yields integral variations in the exchange energy of $\pm4.4\%$ in the exchange regime and $\pm17.6\%$ in the superexchange regime for the respective gate protocols. The sensitivity to lattice imperfections is four times larger compared to the direct exchange gate.

The relative errors for the relevant parameters in our experiment are stated in Extended Data Table \ref{tab:noise_levels}, where the laser intensity fluctuations and inhomogeneity have been directly propagated to the relative tunnelling error, which is about 2.7\% in total. At the same time, band-excitations and motional heating from the non-adiabaticity are negligible, since the modulation frequency is chosen to be much smaller than the band gap. We can estimate the excitation probability with the Landau-Zener formula at the energy gap minimum of the pump cycle: 
\begin{equation}\label{eq:LZ}
    P_e =\exp{\left[-\frac{\pi^2(E_\text{gap}/h)^2}{\partial\nu/\partial\tau}\right]}.
\end{equation}
During the gate operation in our usual lattice configuration the energy gap minimum occurs near the dimerised configuration, with $E_\text{gap}/h\approx 5.6\,$kHz. Averaged over the gate cycle the energy bias is swept at a rate $\partial\nu/\partial\tau=2/h \times \partial\Delta/\partial\tau\approx 21\,$kHz/ms, which yields an excitation probability of $P_e\approx 4\times10^{-7}$. 
Taking the total magnitude of the tunnelling errors together with the on-site interaction fluctuations, this accounts for gate error in the exchange regime of $1.5\%$ and in the superexchange regime of $5.9\%$.
Crucially, the current gate fidelity is limited only by technical noise, such as magnetic field and laser intensity fluctuations.
This contrasts with Rydberg gates, where the finite lifetime of the intermediate state leads to decoherence.
In our gate implementation, further improvements are directly achievable with enhanced experimental stabilisation.

An important difference of our scheme in terms of improving gate fidelity is the fact that a gate operation results purely from an electromagnetic phase shift (Extended Data Fig.~\ref{fig:S4}) which can be simply programmed via an acousto-optic modulator.
In particular, it does not require amplitude modulation or sudden intensity changes of the lattice laser light.
Therefore, the direct exchange gate fidelity can be improved via stabilising the lattice intensity at a fixed lattice depth which lends itself to optimised PID regulation tuning.

{\renewcommand{\arraystretch}{1.2}
\begin{table}[htb]
    \centering
    {\fontfamily{phv}\selectfont 
    \begin{tabular}{lccc}
    \hline\hline
    Noise & Level [\%] & Tunnelling $t$ [\%]  \\
    \hline
         Lattice-depth noise in $V_\mathrm{X}$ & 0.7$^{\text{\tiny{a}}}$  & 1.3\\
         Lattice-depth noise in $V_\mathrm{Xint}$& 2.0$^{\text{\tiny{a}}}$ & 2.1\\
         Lattice-depth noise in $V_\mathrm{Z}$& 0.2$^{\text{\tiny{a}}}$ & 0.2 \\
         Lattice inhomogeneity & 0.6$^{\text{\tiny{b}}}$ & 1 \\
         Fluctuations in Hubbard $U$& 0.8$^{\text{\tiny{a}}}$ & - \\
    \hline
    \end{tabular}
    }
    \caption{\textbf{Noise sources and levels.} $^{\text{\tiny{a}}}$Measured r.m.s.~value. $^{\text{\tiny{b}}}$Computed from independently measured quantities.}
    \label{tab:noise_levels}
\end{table}
}

\begin{figure}[tbp]
    \includegraphics[width=0.48\textwidth]{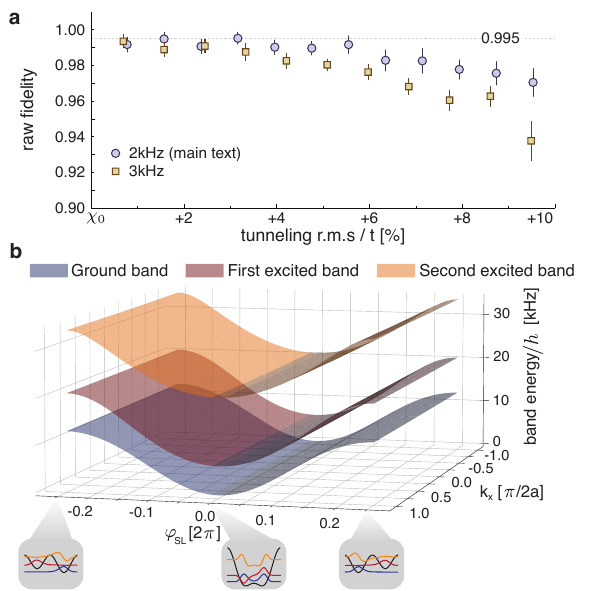}
         \caption{\textbf{Robustness of geometric \swap gate under different noise bandwidths and band structure during the pump cycle.} (a) Raw fidelity as a function of added tunnelling noise for bandwidths of 2 kHz (circles) and 3 kHz (squares). The 2 kHz data is reproduced from Fig.~\ref{fig:3}b for comparison. The 3 kHz data was measured using an identical protocol, with each point averaged over 10 experimental realisations. (b) The calculated energy spectrum of the three lowest Bloch bands as a function of the superlattice phase $\varphi_{\text{SL}}$. The insets show the Wannier functions for the different energy bands in the staggered and dimerised lattice. The geometric \swap gate and bidirectional topological pumping operate within the lowest two bands, which are separated by an energy gap of $5.6\,\text{-}\,8.1$ kHz. The third band is separated by a much larger gap ($>12.4$ kHz), effectively suppressing non-adiabatic excitations.}
    \label{fig:noise+bandstructure}
\end{figure}

The gate's robustness against tunnelling noise (Figs. \ref{fig:3}b and \ref{fig:4}e) depends on the noise bandwidth. For example, increasing the noise bandwidth to 3 kHz shrinks the robustness window to 3\% (Extended Data Fig.~\ref{fig:noise+bandstructure}a). This dependence arises because the gate operation utilizes the lowest two Bloch bands, separated by an energy gap of $5.6\,\text{-}\,8.1$ kHz (Extended Data Fig.~\ref{fig:noise+bandstructure}b). Higher-bandwidth noise can drive non-adiabatic, multi-photon transitions across this gap, and, although much less likely because of the large band separation, it can also induce transitions to higher bands, creating leakage channels that reduce gate fidelity.

\subsection{Realisation with bosons}
\label{subsec:bosons}

In this section, we provide an alternative description of the geometric \swap gate with bosonic particles.
For bosonic systems single-site occupations are not limited to two particles, but for simplicity we focus on the same reduced basis $\{\ket{\uparrow\downarrow,0}, \ket{\uparrow,\downarrow}, \ket{\downarrow,\uparrow},\ket{0,\uparrow\downarrow}\}$ spanning the same Hilbert space as in the fermionic case. The matrix representation is then 
\begin{equation}\label{eq:DW_hamiltonian_bosons_Fock}
\hat{H}_\text{B} = \begin{pmatrix}
U + 2\Delta & -t & -t & 0 \\
-t & 0 & 0 & -t \\
-t & 0 & 0 & -t \\
0 & -t & t & U - 2\Delta
\end{pmatrix}.
\end{equation} 
After a unitary transformation into the ordered basis spanned by $\{\ket{t_0}, \ket{D_+}, \ket{D_-},\ket{s}\}$ the Hamiltonian can be written as 
\begin{equation}
\label{eq:DW_hamiltonian_bosons_STD}
\hat{H}_\text{B} = \begin{pmatrix}
0 & -2t & 0 & 0 \\
-2t & U & 2\Delta & 0 \\
0 & 2\Delta & U & 0 \\
0 & 0 & 0 & 0
\end{pmatrix},
\end{equation} 
where the singlet state $\ket{s}$ remains decoupled, compared to all triplet states being decoupled in the fermionic case. However, the dynamics of states $\ket{t_+}=\ket{\uparrow\uparrow}$ and $\ket{t_-}=\ket{\downarrow\downarrow}$ are given by reduced Hamiltonians of the same form as in equation \eqref{eq:DW_hamiltonian_bosons_STD} due to bosonic statistics in tunnelling amplitudes. The double occupancy states coupled to the $\ket{t_+}$ and $\ket{t_-}$ states are then given by 
\begin{align*}
    \ket{D^\uparrow_\pm} = \frac{1}{\sqrt{2}}(\ket{\uparrow\uparrow,0}\pm\ket{0,\uparrow\uparrow}),\\
    \ket{D^\downarrow_\pm} = \frac{1}{\sqrt{2}}(\ket{\downarrow\downarrow,0}\pm\ket{0,\downarrow\downarrow}).
\end{align*}
Their spectrum is the same as that for $\ket{t_0}$ and all triplet states acquire geometric and dynamical phases during the adiabatic evolution. However, as the scattering length between different internal states varies, dynamical phase contributions for the three triplet states are generally not equivalent. As a result, the control of gate mechanisms in a bosonic system is possible but possesses additional challenges. 

The bosonic Hamiltonian fulfills the same symmetries as the fermionic Hamiltonian discussed in Section \ref{subsec:ExGate_robust}. If $U=0$, the dark state of the bosonic system can be written as 
\begin{equation}
    \ket{\psi} = \cos\left(\frac{\theta}{2}\right)\ket{\mathcal{T}}+\sin\left(\frac{\theta}{2}\right)\ket{D_-}\,,
\end{equation}
for all states of the triplet manifold with their respective $\ket{D_-}$ state. The triplet states are not coupled among each other due to the $\mathrm{SU}(2)$ symmetry of the Hamiltonian. 

\subsection{Data availability}

The data that support the findings of this study are available online at ref.~\cite{kiefer_dataset_2026}.

\end{document}